\documentstyle[aps]{revtex}

\begin{document}
\title{Harmonic scale transformation in effective QFTs with underlying
structures}
\draft
\author{Jifeng Yang}
\address{School of Management, Fudan University, Shanghai 200433, P R China}
\date{August 30 1999}
\maketitle

\begin{abstract}
New equations governing the scale transformation behaviors of a QFT with
underlying structures are derived. These equations, with their several
equivalent versions, can yield some new and significant insights and results
that are difficult to see in the conventional renormalization programs.
Among the several equivalent versions, one is similar to the usual
Callan-Symanzik equation and renormalization group equation but with
different meanings. From another version, the anomalous equation of
energy-momentum tensor trace can be easily obtained. It can be shown that
with the new versions one could partially fix the scheme dependence and
hence some renormalization schemes like momentum space subtractions are
questionable. The asymptotic freedom for QCD is easily reproduced at
one-loop level in our new version CSE. Finally the decoupling theorem a la
Appelquist and Carazzone are discussed within our strategy.
\end{abstract}
\section{Introduction}

According to the standard point of view, the QFTs beset with UV unphysical
divergences could be seen as corresponding to the effective sectors of the
well-defined and complete underlying theory without UV (as well as IR)
troubles. The present QFTs are just ill-defined formulations of the
well-defined sectors in the underlying theory. From this point of view we
proposed a natural strategy for calculating radiative corrections without
introducing ad hoc regularizations and UV divergences \cite{Usa}.

In this report, we will derive the equations governing the scale
transformation behavior of the QFTs with the existence of underlying
structures for making the UV (to focus on the main topic we assume from now
on the IR ends are IR regular) ends of the effective QFTs well-defined. Our
derivation does not require anything about the renormalizability as will be
clear below. Thus the equations obtained are valid for all QFTs as long as
they are consistent and/or physically relevant. For the renormalizable
cases, the new equations are just new versions of the conventional
Callan-Symanzik equations \cite{CS}.

This paper is organized in the following way: In section two we derive the
equations governing the scale behavior of an arbitrary QFT with underlying
structures to make it UV (and IR) well defined. The Callan-Symanzik (CS)
equation is given in several equivalent versions in section three. The
anomalous trace equation of the energy-momentum tensor follows naturally. In
section four, we will elaborate on the harmony revealed by the new scale WT
identities and its implications for the conventional renormalization
schemes. The scheme dependence of the decoupling theorem \`{a} la Appelquist
and Carazzone \cite{Appel} is discussed in section five in the underlying
theory point of view. In section six we summarize our presentation.

\section{Underlying structures and scale anomalies}

Before starting the derivation, we briefly review our strategy and its
consequences: Suppose that the complete and well-defined underlying theory
is found, then it must contain certain additional parameters or constants
characterizing the correct UV structures underlying the present QFTs. The
latter can be now formulated without any (UV) ill-definedness in the
underlying theory. As the underlying structures are too small comparing to
the scales of the effective theories, the latter are effectively defined in
the limit that the underlying structures vanish. But this does not mean that
the underlying structures are totally decoupled. On the contrary, as pointed
out in Ref. \cite{Usa}, the underlying structures do ''influence'' the
effective theories through certain agent constants that are not included in
the present QFT formulations. The well-defined formulation of the QFTs are
in fact defined in terms of by the canonical model constants of QFTs AND the
agent constants. There is no room in our proposal for divergence and counter
terms. Hence the meaning of all the finite constants in our proposal is
different from the conventionally 'renormalized' ones.

Now let us begin our derivation. Consider a general complete vertex function
(or a 1PI Green function), $\Gamma^{(n)}([p],[g];\{\sigma\})$ given in the
underlying theory with $[p],[g]$ denoting the external momenta and the
Lagrangian couplings (including masses) and $\{\sigma\}$ denoting the
underlying parameters or constants. Now it is easy to see that any
well-defined vertex function {\bf must} be a homogeneous function of all its
dimensional arguments, that is 
\FL
\begin{eqnarray}
\Gamma^{(n)}([\lambda p],[\lambda^{d_g} g]; \{ \lambda^{d_{\sigma}}
\sigma\})= \lambda^{d_{\Gamma^{(n)}}} \Gamma^{(n)} ([p],[g];\{\sigma\})
\end{eqnarray}
where $d_{\cdots}$ refers to the canonical scale dimension of a parameter or
canonical mass dimension of a constant (effective and/or underlying).

In the limit ($L_{\{\sigma \}}$) that the underlying structures vanish,
there will necessarily arise some well-defined agent constants $\{\bar{c}\}$
\cite{Usa}, then Eq.(1) becomes 
\FL
\begin{eqnarray}
&&\Gamma ^{(n)}([\lambda p],[\lambda ^{d_g}g];\{\lambda
^{d_{\bar{c}}}{\bar{c%
}}\})\equiv L_{\{\sigma \}}\Gamma ^{(n)}([\lambda p],[\lambda
^{d_g}g];\{\lambda ^{d_\sigma }\sigma \})  \nonumber \\
&=&\lambda ^{d_{\Gamma ^{(n)}}}\Gamma ^{(n)}([p],[g];\{{\bar{c}}\})\equiv
\lambda ^{d_{\Gamma ^{(n)}}}L_{\{\sigma \}}\Gamma ^{(n)}([p],[g];\{\sigma
\}).
\end{eqnarray}
where it is understood that even the smallest mass scale of the underlying
constants is infinitely large for all the scales appearing in the effective
sectors (masses, couplings or typical energy scales). Note that these
constants only appear in the loop diagrams of certain vertex functions or in
the pure quantum corrections. The limit operation will lead to a local part
of certain vertex functions with their coefficients as functions of the
agent constants $\bar{c}$ and the Lagrangian constants \footnote{%
Here the notation $\bar c$ takes a different meaning in contrast to Ref.
\cite{Usa} where we use $\bar c$ to denote just the definite coefficients of
the local part of a vertex function. Later we will denote the coefficients
of the local vertices as $\{c\}$.}. In all the following deductions, it is
understood that for any equation derived in the underlying version ($\Gamma
(\cdot ;\{\sigma \})$) there is a parallel version defined in terms of the
agent constants $\{{\bar{c}}\}$. Let us work in the underlying version
first.

Now the differential equation version of Eq. (1) reads 
\FL
\begin{eqnarray}
\{\sum_{i=1}^n p_i \cdot \partial_{p_i} + \sum d_g g\partial_g + \sum
d_{\sigma} \sigma \partial_{\sigma} -d_{\Gamma^{(n)}}\}
\Gamma^{(n)}([p],[g];\{\sigma\})=0.
\end{eqnarray}
All the scale or mass dimensions are canonical here. Obviously the constants
with zero canonical mass dimensions do not cause any violation of the scale
behavior. Eqs. (1) to (3) are valid for any QFT provided they are
mathematically consistent and/or physically effective. Renormalizability is
simply not in concern here. It is clear that the scale anomalies are caused
by the indispensable existence of the underlying structures.

Note that the operation $d_g g\partial_g$ in the above equations inserts
certain local operators into the vertex function under consideration, which
is in turn equal to the trace of the energy-momentum tensor ($\Theta$) for
the QFTs in concern (it is an easy exercise to check this). Thus we arrive
at the following, 
\FL
\begin{eqnarray}
\{\sum_{i=1}^n p_i \cdot \partial_{p_i} + \sum d_{\sigma} \sigma
\partial_{\sigma} -d_{\Gamma^{(n)}}\} \Gamma^{(n)}([p],[g];\{\sigma\})
=-i\Gamma^{(n)}_{\Theta} ([0;p],[g];\{\sigma\})
\end{eqnarray}

Eq.(3) or equivalently Eq.(4) is just the most general underlying structure
version of the Ward-Takahashi identity for scale transformation in
(effective) QFTs. The anomalies in the effective QFTs are {\bf canonical
behaviors} in the underlying theory point of view as the underlying
constants $\{\sigma\}$ are featuring constants there. The parallel equation
of Eq. (3) in terms of the agent constants $\{\bar c\}$ that can be derived
from Eq.(2) read 
\FL
\begin{eqnarray}
\{ \sum_{i=1}^n p_i \cdot \partial_{p_i} + \sum d_{\bar c} {\bar c}
\partial_{\bar c} +\sum d_g g \partial_g -d_{\Gamma^{(n)}} \}
\Gamma^{(n)}([p],[g];\{\bar c\})=0.
\end{eqnarray}
Eqs. (3), (4) and (5) are valid not only order by order but also graph by
graph. It is very clear that the agent constants, which arise obviously from
the limit operation $L_{\{\sigma\}}$, take over all the responsibilities
originally held by the underlying structures $\{\sigma\}$. {\bf They are
inherent in the well-defined forms of the QFTs. They reside not in the tree
graph section of the theories but only in the pure quantum loop graphs}.

Now we note the following obvious fact: Since the agent constants are
arguments of the coefficients ($c_{O;\gamma }([g];\{\bar{c}\})$) of the
local part of certain vertex functions' loop graph (denoted as $\gamma $)
due to the limit operation, the variations in these agent constants finally
lead to the insertions of the local vertex operators ($[O]$). Some of these
composite vertex operators are just given by the Lagrangian. Then the
insertion of some of these operators can be realized through the variation
with respect to the couplings in the Lagrangian (the insertion of the
kinetic terms and others that can not be found in the Lagrangians will be
generally denoted as ${\hat{I}}_{O_N}$). That is, with the following facts
valid for the complete vertex functions, 
\FL
\begin{eqnarray}
&&\delta _{O;\gamma }([g];\{\bar{c}\})\equiv
D_{c_{O;\gamma }}([g];\{\bar{c}%
\})\equiv \sum_{\{\bar{c}\}}d_{\bar{c}}\frac{{\bar{c}}\partial
c_{O;\gamma }%
}{c_{O;\gamma }\partial \bar{c}}, \\
&&\sum d_{\bar{c}}{\bar{c}}\partial _{\bar{c}}=\sum_{\{O;\gamma
\}}D_{c_{O;\gamma }}c_{O;\gamma }\partial _{c_{O;\gamma
}}=\sum_{\{c_L\}}D_{c_L}c_L\partial _{c_L}+\sum_{\{c_N\}}D_{c_N}c_N\partial
_{c_N},  \nonumber \\
&& \\
&&\sum_{\{c_L\}}D_{c_L}c_L\partial _{c_L}=\sum_{[g]}\delta _g([g];\{\bar{c}%
\})g\partial _g+\sum_{\{\phi \}}\delta _\phi ([g];\{\bar{c}\}){\hat{I}}_\phi
, \\
&&\sum_{\{c_N\}}D_{c_N}c_N\partial _{c_N}=\sum_{\{O_N\}}\delta
_{O_N}([g];\{%
\bar{c}\}){\hat{I}}_{O_N},
\end{eqnarray}
we can turn Eq.(5) into the following forms, 
\FL
\begin{eqnarray}
&&\{\lambda \partial _\lambda +\sum_{\{c_{O;\gamma }\}}D_{c_{O;\gamma
}}c_{O;\gamma }\partial _{c_{O;\gamma }}+\sum_{[g]}d_gg\partial _g-d_{\Gamma
^{(n)}}\}\Gamma ^{(n)}([\lambda p],[g]|\{c_{O;\gamma }\})=0, \\
&&\{\lambda \partial _\lambda +\sum_{\{O_N\}}\delta _{O_N}{\hat{I}}%
_{O_N}+\sum_{[g]}(\delta _g+d_g)g\partial _g+\sum_{\{\phi \}}\delta _\phi {%
\hat{I}}_\phi -d_{\Gamma ^{(n)}}\}\Gamma ^{(n)}([\lambda
p],[g]|\{c_{O;\gamma }\})=0,
\end{eqnarray}
Here the insertion of the kinetic vertex of a field $\phi $ is denoted as
${%
\hat{I}}_\phi $. $\delta _{\cdots }$ or $D_{c_{\cdots }}$ denote the
coefficients before the inserted composite operators. (A vertex function's
dependence upon the agent constants is effected through the local
coefficient constants after the vertical bar.) Eq.(10) is true order by
order, graph by graph while Eq.(11) is only true for the complete sum of all
graphs (or sum up to a certain order). The constants $\{c_{\cdots }\}$ must
satisfy all the above equations as functions of the agent constants {\bf
and}
the model constants that are classically defined. Thus one can view the
above equations as constraints imposed by the most natural
fact--homogeneity--upon these constants. Now the anomalies are defined by
the $\delta$ functions in front of the insertion operators.

Although in Eq.(11) we have rewritten some of anomalies in terms of the
insertions of the composite operators appearing in the Lagrangian, they are
specifying the scale behaviors of the theories's {\bf quantum structures},
not that of the classical constants. In fact they describe the influences of
the "normal" scale behaviors of the underlying structures upon the effective
QFTs. Later, in section four, we will discuss more closely about this point
followed by some significant consequences which have not been achieved in
the conventional renormalization programs.

\section{New versions of Callan-Symanzik equation and RGE}

In this section we limit our attention to a special type of theories where
there is no variations in certain local vertices (those denoted by $\sum
D_{c_N}c_N \partial_{c_N}$ or $\sum_{\{O_N\}} \delta_{O_N}([g];\{\bar c\})
{%
\hat{I}}_{O_N}$ in Eq.(9)) that can not be realized through insertions of
the Lagrangian operators. In conventional terminology, we consider in this
section only renormalizable theories.

Without the variations necessarily described by $\sum D_{c_N}c_N\partial
_{c_N}$, Eqs. (10) and (11) should take the following form, 
\FL
\begin{eqnarray}
\{\lambda \partial _\lambda +\sum d_gg\partial
_g+\sum_{\{c_L\}}D_{c_L}c_L\partial _{c_L}-d_{\Gamma ^{(n)}}\}\Gamma
^{(n)}([\lambda p],[g]|\{c_{\cdots }\})=0, \\
\{\lambda \partial _\lambda +\sum (\delta _g+d_g)g\partial _g+\sum \delta
_\phi {\hat{I}}_\phi -d_{\Gamma ^{(n)}}\}\Gamma ^{(n)}([\lambda
p],[g]|\{c_{\cdots }\})=0,
\end{eqnarray}
These are just some new versions of the CS equations for renormalizable
theories. Before turning them into more familiar forms,let us first make the
following remarks.

As pointed out in section two, the most striking difference between $[g]$
and $\{\bar c\}$ is that the former are defined in the {\bf tree level
vertices} while the latter only appear in {\bf loop amplitudes}. The
classical constants $[g]$ or the tree level vertices alone can not yield a
well-defined definition of the quantum structures of fields for theories
with potential UV divergence. But the incompleteness of the constants $[g]$
or the unavailability of $\{\sigma\}$ by now does not mean that there must
be divergence. Of course among $\{\sigma\}$ or their agents $\{\bar c\}$
that characterize the quantum "paths", there must be ones with mass
dimensions specifying the effective quantum motions' scale(s). In our
opinion, this is the physical basis of the necessary appearance of a mass or
energy scale in any regularization/renormalization scheme. In our strategy,
these scales naturally appear in the course of indefinite integrations.

On the other hand, the locality of the appearance of these quantum agents
does allow a mix among $[g]$ and the agent constants (through $%
\{c_{\gamma}([g];\{\bar c\})\}$) {\bf for the complete} vertex functions or
their generating functional in renormalizable theories. Suppose there are
scale transforms in the agent constants, then 
\FL
\begin{eqnarray}
& &\sum_{\{\bar c\}} d_{\bar c} {\bar c} \partial_{\bar c} \Gamma^{(n)}
(\cdots|\{c_{\cdots} ([g];\{\bar c\})\})= \sum_{\{c_{\cdots}\}}
D_{c_{\cdots}} c_{\cdots} \partial_{c_{\cdots}} \Gamma^{(n)}
(\cdots|\{c_{\cdots} ([g];\{\bar c\})\})  \nonumber \\
&=&\{ \sum_{[g]} \delta_g g\partial_g +\sum_{\{\phi\}} \delta_{\phi}
{\hat{I}%
}_{\phi}\}\Gamma^{(n)} (\cdots| \{c_{\cdots} ([g];\{\bar c\})\})
\end{eqnarray}
which is equivalent to 
\FL
\begin{eqnarray}
& &\{\sum_{\{\bar c\}}d_{\bar c}{\bar c}\partial_{\bar c}
-\sum_{\{c_{\cdots}\}} D_{c_{\cdots}} c_{\cdots} \partial_{c_{\cdots}}
\}\Gamma^{(n)} (\cdots|\{c_{\cdots} ([g];\{\bar c\})\})  \nonumber \\
&=&\{\sum_{\{\bar c\}}d_{\bar c}{\bar c}\partial_{\bar c} - \sum_{[g]}
\delta_g g\partial_g -\sum_{\{\phi\}} \delta_{\phi} {\hat{I}}%
_{\phi}\}\Gamma^{(n)} (\cdots|\{c_{\cdots} ([g];\{\bar c\})\})=0.
\end{eqnarray}
It is easy to see that the "invariance" in Eq.(15) is in fact a
rearrangement of the classically defined parts and that arising from quantum
loops or a tautology of the fact that agent constants' changes only affect
the elementary like local vertices as evident in Eq.(14). This is our new
version of the renormalization group equation.

Now let us turn the kinetic terms' contribution into other forms. This can
be achieved by noting that the coefficient constants in the kinetic vertices
only "renormalize" or rescale the line momenta in the graphs. For example,
for the fermions, this is just the rescaling of $ip\!\!\!/ \rightarrow
(1+c_{\psi})i p\!\!\!/$. Similarly for bosons, $ip^2 \rightarrow
(1+c_{\phi})ip^2$. Since the lines must end up in various vertices, such
rescaling of lines would effectively lead to the rescaling of vertices and
be shared by the latter in pairs. Thus a vertex is typically rescaled as $%
\prod_i (1+c_{\psi_i})^{-1/2}\prod_j(1+c_{\phi_j})^{-1/2}$ depending on the
number of line momenta that join it. For an $n$-point 1PI vertex function
there must be $n$ external momenta flow into or out of a number of
elementary vertices in the graphs. As the external momenta are not subject
to the rescaling there must be compensated rescaling for these vertices
containing external momenta--an overall rescaling of the complete $n$-point
vertex functions. Thus we have 
\FL
\begin{eqnarray}
& &\delta_g \rightarrow {\bar \delta}_g\equiv (\delta_g- n_{g;\phi}\frac{%
\delta_{\phi}}{2}-n_{g;\psi} \frac{\delta_{\psi}}{2}), \ \ d_g=0; \delta_g
\rightarrow {\bar \delta}_g\equiv (\delta_g- \delta_{\psi or \phi}), \ \
d_g\neq 0;  \nonumber \\
& &\Gamma^{(n_{\phi},n_{\psi})} \rightarrow
(1+c_{\psi})^{n_{\psi}/2}(1+c_{\phi})^{n_{\phi}/2}
\Gamma^{(n_{\phi},n_{\psi})}.
\end{eqnarray}
Then the above equations take the following forms: 
\FL
\begin{eqnarray}
&&\{ \lambda \partial_{\lambda}+ \sum ({\bar \delta}_g +d_g)g \partial_g +
n_{\phi}\frac{\delta_{\phi}}{2} + n_{\psi}\frac{\delta_{\psi}}{2}%
-d_{\Gamma^{(n_{\phi},n_{\psi})}}\} \Gamma^{(n_{\phi},n_{\psi})}
([\lambda p],[g]|\{c_{\cdots}\}) =0, \\
&&\{\sum_{\{\bar c\}}d_{\bar c}{\bar c}\partial_{\bar c} - \sum_{[g]} {\bar
\delta}_g g\partial_g -n_{\phi} \frac{\delta_{\phi}}{2}-n_{\psi}\frac{%
\delta_{\psi}}{2} \} \Gamma^{(n_{\phi},n_{\psi})}
([p],[g]|\{c_{\cdots}\})=0.
\end{eqnarray}
One can see that Eqs.(17) and (18) take exactly the same form as the usual
CSE and RGE. But we must emphasize again here that everything here is given
in terms of the classical constants $[g]$ and agent constants. One might
take the $[g]$ as our finite "bare" constants.

Now some remarks are in order:

(1). Each equation is given in terms of the classically defined constants $%
[g]$ and the quantum agents definitely defined in the complete underlying
theory, no divergence and hence no infinite renormalization is needed. In
the conventional schemes, divergence is inevitable due to the artificiality
of the regularization methods and hence infinite renormalization is
necessary. Then after subtraction, one needs to {\bf "reidentify"} the
originally classically defined constants with a product of two singular
objects--bare quantities and the renormalization constants.

(2). Conventionally, the infinite bare quantities are in fact regularization
scheme dependent--different divergence in different schemes, therefore
scheme dependence is inherent there as infinite renormalization only
transforms the expressions not the essence. In the underlying theory point
of view this problem is due to the artificiality of the regularization
schemes for simulating the true underlying structures \footnote{%
Conventionally, it is believed that the scheme dependence is only inherent
in the perturbative truncations but not in the full renormalized theory,
basing on the expectation that the low energy theories are independent of
the short distance structures and hence independent of regularization
schemes. This is not correct as even the underlying constants do not show up
in the effective theory, they do affect the effective theories through their
agent constants. The scale anomalies are just the evidences of the
influences of the underlying structures. Otherwise, why bother choosing
among regularizations?}. There should be only one physical scheme, the one
with the true agent constants selected by the nature together with the
classically defined Lagrangian constants apart from some redefinition
freedom for the special type theories described in Eqs. (14), (15) or (18).

Let us take QED as an example for demonstration.

The CS equation and RGE in conventional renormalization schemes read
respectively \cite{Love} 
\FL
\begin{eqnarray}
&&\{ \lambda \partial_{\lambda}+(1+\gamma_{m_R})m_R \partial_{m_R}-\beta
\partial_{\alpha_R} +n_A\gamma_A -n_{\psi}\gamma_{\psi}
+d_{\Gamma^{(n_A,n_{\psi})}} \} \Gamma^{(n_A,n_{\psi})}([\lambda p],m_R,
\alpha_R)=0; \\
&&\{\mu\partial_{\mu} -\gamma_{m_R} m_R \partial_{m_R} +\beta
\partial_{\alpha_R}-n_A\gamma_A
-n_{\psi}\gamma_{\psi}\}\Gamma^{(n_A,n_{\psi})} ([p], m_R, \alpha_R)=0.
\end{eqnarray}
While to write down our new version CSE and RGE for QED we need to be
specific on the following aspects: Among the agent constants there is at
least one mass scale parameter, we will denote it as $\bar \mu$ and
parametrize all the agents into one mass scale and a series of constants
$\{%
\bar c_{i;0}\}$ with zero mass dimension so that the coefficients take the
following form 
\FL
\begin{equation}
c_m = C_m(m,e;{\bar \mu})+C_{m;0}(\{{\bar c}_{i;0}\}),\ \ c_e = C_e(m,e;{%
\bar \mu})+C_{e;0}(\{{\bar c}_{i;0}\})= c_{\psi},\ldots.
\end{equation}
where the Ward-Takahashi identity for gauge invariance is used. Then from
Eqs.(12) to (18) we can write down our new CSE and RGE in several equivalent
versions as follows 
\FL
\begin{eqnarray}
&&\{\lambda\partial_{\lambda} + m\partial_m +D_{c_m}c_m\partial_{c_m}
+D_{c_A}c_A\partial_{c_A}+ D_{c_{\psi}}c_{\psi}\partial_{c_{\psi}}
-d_{\Gamma^{(n_A,n_{\psi})}}\} \Gamma^{(n_A,n_{\psi})} ([\lambda p],
m,e|\{c_m,c_e,c_A,c_{\psi}\})  \nonumber \\
&&=\{\lambda\partial_{\lambda}+(1+\delta_m)m\partial_m +\delta_e e\partial_e
+\delta_A {\hat{I}}_A+\delta_{\psi} {\hat{I}}_{\psi}
-d_{\Gamma^{(n_A,n_{\psi})}}\}\Gamma^{(n_A,n_{\psi})} ([\lambda p], m,e|
\{c_m,c_e,c_A,c_{\psi}\})  \nonumber \\
&&=\{ \lambda \partial_{\lambda}+ ({\bar \delta}_m +1)m \partial_m +\frac{{%
\bar \delta}_e}{2}e\partial_e+ n_A\frac{\delta_A}{2} + n_{\psi}\frac{%
\delta_{\psi}}{2} -d_{\Gamma^{(n_A,n_{\psi})}}\} \Gamma^{(n)}([\lambda p],
m,e| \{c_m,c_e,c_A,c_{\psi}\}) =0; \\
&&\{{\bar \mu} \partial_{\bar \mu} -D_{c_m}c_m\partial_{c_m}
-D_{c_A}c_A\partial_{c_A}-D_{c_{\psi}}c_{\psi}\partial_{c_{\psi}}\}
\Gamma^{(n_A,n_{\psi})}([p], m,e| \{c_m,c_e,c_A,c_{\psi}\})  \nonumber \\
&&=\{{\bar \mu} \partial_{\bar \mu} -\delta_m m\partial_m-\delta_e
e\partial_e-\delta_A {\hat{I}}_A -\delta_{\psi} {\hat{I}}_{\psi}\}
\Gamma^{(n_A,n_{\psi})} ([p], m,e|\{c_m,c_e,c_A,c_{\psi}\})  \nonumber \\
&&=\{ {\bar \mu} \partial_{\bar \mu} - {\bar \delta}_m m\partial_m -\frac{{\
\bar \delta}_e}{2} e\partial_e - n_{\phi}\frac{\delta_{\phi}}{2}- n_{\psi}
\frac{\delta_{\psi}}{2}\} \Gamma^{(n_A,n_{\psi})} ([p],
m,e|\{c_m,c_e,c_A,c_{\psi}\})=0.
\end{eqnarray}

Now we can compare our new CSE and RGE in (22) and (23) with the
conventional ones in (19) and (20). The anomalies are all described in our
equations in terms of the classical constants $m,e$ or ($\alpha$) and the
agents ${\bar \mu}, {\bar c_{\cdots}}$, while in the conventional
renormalization schemes the anomalies are given in terms of the renormalized
constants $m_R, \alpha_R$ depending on an arbitrary scale $\mu$ via the
infinite renormalization constants. To the lowest order the correspondence
should be 
\FL
\begin{eqnarray}
\gamma_{m_R} \sim {\bar \delta}_m\equiv (\delta_m-\delta_{\psi}) ;\ \ \beta
(\alpha_R)=2\gamma_A \sim \delta_A \sim -2{\bar \delta}_e;\ \ 2\gamma_{\psi}
\sim \delta_{\psi}.
\end{eqnarray}
We note again that the equations expressed in terms of the coefficient
constants directly are valid order by order and graph by graph as well as
for the complete sum or sum up to a certain order while the ones given in
terms of vertex insertions are only valid for the complete sum or sum up to
certain order (so it is with the conventional CSE and RGE).

From the experiences above it is also easy to derive the WT identity of
scale transformation for generating functional of vertex functions of a QFT
from homogeneity, 
\FL
\begin{eqnarray}
&&\{\sum_{\{\phi\}}\int d^D x [d_{\phi}-x\cdot\partial_x) \phi(x)] \frac {%
\delta}{\delta \phi(x)} + \sum_{[g]}d_gg\partial_g +\sum_{\{\bar c\}} d_{%
\bar c} {\bar c} \partial_{\bar c}-D\} \Gamma^{1PI} ([\phi],[g];\{\bar c\})
\nonumber \\
&&=\{\sum_{\{\phi\}}\int d^D x [d_{\phi}-x\cdot\partial_x) \phi(x)] \frac {%
\delta}{\delta \phi(x)} + \sum_{[g]}(d_g+\delta_g)\partial_g
+\sum_{\{\phi\}} \delta_{\phi}{\hat{I}}_{\phi}-D\} \Gamma^{1PI}
([\phi],[g];\{\bar c\}) =0.
\end{eqnarray}
with $D$ denoting the spacetime dimension. Again this is correct for any
consistent QFTs.

To exhibit the utility of the version in terms of the operator insertion, we
simply note that once it is translated into the operator form we get the
anomalous trace equation of quantum energy-momentum tensor since all the
quantum effects all well defined here and all the effects have been given in
terms of known quantum operators. Thus one can just read the trace equation
from the operator insertion version of Eq.(25), for QED this is simply
\FL
\begin{eqnarray}
g_{\mu \nu }{\Theta }^{\mu \nu }=(1+\delta _m)m{\bar{\psi}}\psi +\frac 14%
\delta _AF^{\mu \nu }F_{\mu \nu }-\delta _\psi i{\bar{\psi}}\gamma _\mu
D^\mu \psi =(1+\delta _m-\delta _\psi )m{\bar{\psi}}\psi +\frac 14\delta
_AF^{\mu \nu }F_{\mu \nu }.
\end{eqnarray}
In the last step we have used the motion equation. Considering the
correspondence in Eq. (24), this is exactly the operator trace anomaly
equation for QED \cite{Trace} in a new version without divergence and
subtractions. In unrenormalizable theories, the trace anomalies would
contain an infinite sum of local composite operators beyond the Lagrangian
operators.

In next section we will illustrate the correctness of our new WT identity or
new CS equation in unconventional versions in several examples and
demonstrate their physical utility and significance.

\section{Harmony constraint on the agent constants}

We have seen that vertex functions or their generating functional for a QFT
must be homogeneous functions with respect to {\bf all} the dimensional
parameters and constants which must include the underlying structures or
their agents. That is to say, for this {\bf homogeneity} to be valid in a
well-defined formulation the underlying structures are indispensable.
Therefore in the underlying theory point of view the scale transform is in
perfect harmony in the following sense: When all the arguments scale up
together according to their normal scale or mass dimensions the vertex
functions(functional) must then exhibit exact scale behaviors. This is the
harmony principle we wish to advocate that is understandable only in terms
of underlying structures.

This harmony leads to a natural explanation for the scale or trace anomaly.
On the one hand, in the underlying theory formulation, there is no scale
anomaly. The scale behavior is definite and normal there. In the rescaling
of all fields and all constants, there should be generally transformations
into each other (just like any symmetry transformation) in a definite way to
preserve the overall homogeneity. Then as we only effectively observe the
relatively low sectors' variables and their behavior, the anomalous scale
variations of the effective variables that should be absorbed by the
underlying structures' variations are therefore exposed. Thus it is equally
correct to interpret the anomalies as the normal contributions of underlying
structures and as the anomalous behaviors of the effective sectors that
should be unambiguous in terms of the canonical variables of effective
theories \footnote{%
We had shown elsewhere \cite{NTrace} and will further demonstrate below that
scale or trace anomaly is in fact due to the presence of a kind of rational
terms that are nonlocal (hence definite at least in 1-loop level). The
presence of this kind of rational terms is unambiguous or independent of
regularization schemes. It is in this sense the anomaly is a definite
property of QFT.}. They are the two faces of one thing. Chiral anomaly can
be understood in similar way \cite{Hooft}. We hope this harmony principle
could be generalized to other transformation behaviors of the effective
theories.

It is known in gauge field theories that gauge invariance reduces the number
of ambiguities (or the number of independent divergences in conventional
terminology). Here, with the harmony notion elaborated above, further
important constraints upon the ambiguities can be imposed. This scale
harmony has in fact locked the scale behaviors of the agent constants with
that of the definite parts of the effective theories and hence the
ambiguities are further reduced. Now let us enumerate some examples to
illustrate the validity of our new CS equations and the role of ambiguity
reduction of these equations in the notion of harmony.

Let us illustrate our point of view with simple examples, i.e., the 1-loop
level elementary vertices in QED: electron self-energy $\Sigma ^{(1)}$,
electron-photon vertex $\Lambda _\mu ^{(1)}$ and photon polarization $\Pi
^{\mu \nu }$. For simplicity let us consider the cases of massless
electrons. As for the expressions of these objects in terms of the unknown
agent constants, one can use any regularization to compute the definite part
and just leave the ambiguous local part as an ambiguous polynomial of
external momentum. Or one can use the technique introduced in Ref.
\cite{Usa}
. (We will also explain this technique later.) It is sufficient to
parametrize the agent constants with one dimensional constant ${\bar{\mu}}$
and a number of dimensionless constants. Then from homogeneity we find that
\FL
\begin{eqnarray}
&&\Sigma ^{(1)}(p,-p)=-i\frac{e^2}{16\pi ^2}p\!\!\!/[\ln
\frac{-p^2}{{\bar{\mu}%
}^2}+c_\psi ^0(\{{\bar{c}}^0\})];\nonumber \\
&&\Lambda _\nu ^{(1)}(p,0|p)=+i\frac{e^2}{16\pi ^2}\gamma _\nu [\ln
\frac{-p^2%
}{{\bar{\mu}}^2}+2p\!\!\!/p_\nu /p^2+c_e^0(\{{\bar{c}}^0\})];
\nonumber \\
&& \Pi ^{(1)\mu \nu }(p,-p)=i\frac{e^2}{12\pi
^2}(p^2g^{\mu \nu }-p^\mu p^\nu )[\ln
\frac{-p^2}{{\bar{\mu}}^2}+c_A^0(\{{\bar{c}}^0\})].
\end{eqnarray}
Here all the constants are meant for 1-loop case and the superscript $0$ of
the constants means that these constants are dimensionless and independent
of any dimensional constants and we work in Feynman gauge. Then these
objects have scale anomalies in terms of canonical variables only (here only
momentum) 
\FL
\begin{equation}
\{p\cdot \partial _p-1\}\Sigma ^{(1)}=-i\frac{e^2}{8\pi ^2}p\!\!\!/; \ \ \
\{p\cdot \partial _p\}\Lambda _\nu ^{(1)}=+i\frac{e^2}{8\pi ^2}\gamma _\nu ;
\ \ \ \{p\cdot \partial _p-2\}\Pi ^{(1)\mu \nu }=i\frac{e^2}{6\pi ^2}%
(p^2g^{\mu \nu }-p^\mu p^\nu ).
\end{equation}

Now if we include the agent constants' contributions, which is ${\bar{\mu}}%
\partial _{\bar{\mu}}$ (the dimensionless agent simply do not affect scale
behavior), then we can recover the exact scale behavior or the exact scale
harmony in the version with explicit agent constants which is valid order by
order, graph by graph: 
\FL
\begin{equation}
\{p\cdot \partial _p+{\bar{\mu}}\partial _{\bar{\mu}}-1\}\Sigma ^{(1)}=0; \
\ \ \ \{p\cdot \partial _p+{\bar{\mu}}\partial _{\bar{\mu}}\}\Lambda _\nu
^{(1)}=0; \ \ \ \ \{p\cdot\partial _p+{\bar{\mu}}\partial _{\bar{\mu}}-2\}
\Pi^{(1)\mu \nu }=0.
\end{equation}
This is just harmony we have explained above. The dimensional agent constant
${\bar{\mu}}$ must appear in the vertices as $\ln {\bar{\mu}}^2$. They must
compensate the unambiguous anomalies from the definite and nonlocal parts.
(But the other dimensionless agent constants remain undetermined by the
Scale WT identities.) It is easy to see that the anomaly in each 1-loop
vertex above is due to the logarithmic dependence upon momentum which upon
the action of $p\cdot \partial _p$ yields a local term from a nonlocal one:
$%
(p\cdot \partial _p)\ln[- p^2]=p_\mu (\frac{2p^\mu }{p^2})=2$. Or if one
starts with the insertion of the dilatation current into the elementary
vertices there should be an unambiguous rational term in terms of momentum
only like $\sim \frac{p^\mu p_{\cdots }}{p^2}$ which yields a local terms
after contracting with divergence operation $p_\mu \times \frac{p^\mu
p_{\cdots }}{p^2}=p_{\cdots }$. By the way, we note that the conventional
version of CSE and the one in terms of vertex insertions can not hold order
by order and graph by graph but only for the complete vertex functions or
sums up to certain order.

If one sums the vertices up to 1-loop level with tree graph vertices
included she/he could get the vertex insertion version as follows, 
\FL
\begin{eqnarray}
&&\{p\cdot\partial_p+\delta_{\psi}{\hat{I}}_{\psi}-1\}
\Gamma^{(1)}_{\psi}=0;\ \ \ \ \{p\cdot\partial_p+\delta_e e\partial_e\}
\Gamma^{(1)}_{e;\mu}=0; \ \ \ \ \{p\cdot\partial_p+\delta_A {\hat{I}}_A-2\}
\Gamma^{(1)\mu\nu}=0; \\
&&\delta^{(1)}_{\psi}=\frac{e^2}{8\pi^2};\ \ \ \delta^{(1)}_e =\frac{e^2}{
8\pi^2};\ \ \ \delta^{(1)}_A =\frac{e^2}{6\pi^2}.
\end{eqnarray}

It is easy to see that the anomalies are independent of mass. One can
recalculate the 1-loop corrections of the above vertices in the massive case
with the technique we proposed \cite{Usa}. Here we just list the photon
polarization case given by Chanowitz and Ellis \cite{Ell} but parametrize
the local parts following our strategy,
\FL
\begin{equation}
\Pi ^{\mu \nu }=-i\frac{e^2}{12\pi ^2}(p^\mu p^\nu -g^{\mu \nu
}p^2)\{c_A(m;\{{\bar{\mu}},{\bar{c}}_i^0\})+6\int_0^1dzz(1-z)\ln
[1-z(1-z)p^2/m^2]\};
\end{equation}
\FL
\begin{equation}
\Delta ^{\mu \nu }=\frac{e^2}{\pi ^2}(p^\mu p^\nu -g^{\mu
\nu }p^2)\frac{m^2%
}{p^2}(m^2 F-1),\ \ \ F\equiv \frac 2{\sqrt{(p^4-4m^2p^2)}}\ln
\frac{p^2-\sqrt{%
(p^4-4m^2p^2)}}{p^2+\sqrt{(p^4-4m^2p^2)}},
\end{equation}
\FL
\begin{equation}
\{p\cdot \partial _p-2\}\Pi ^{\mu \nu }=i\Delta ^{\mu \nu }+i\frac{e^2}{6\pi
^2}(g^{\mu \nu }p^2-p^\mu p^\nu )
\end{equation}
where the anomaly is exactly the same as in the massless case.

Now if we include the ${\bar{\mu}}\partial _{\bar{\mu}}$ operation and that
of the mass, there should not be any anomaly as all dimensional arguments
are harmoniously considered. Thus the anomaly must be assumed up by the
dimensional agent's variation, or the dimensional agent must be locked in
this way with the definite anomaly, 
\FL
\begin{eqnarray}
&&{\bar{\mu}}\partial _{\bar{\mu}}c_A(m;\{{\bar{\mu}},{\bar{c}}_i^0\})=-2 \\
&&\longrightarrow c_A(m;\{{\bar{\mu}},{\bar{c}}_i^0\})=-\ln
\frac{{\bar{\mu}}%
^2}{m^2}+c_A(\{{\bar{c}}_i^0\})
\end{eqnarray}
where the only dimensional constant besides the agent ${\bar{\mu}}$ is
fermion mass here \footnote{%
One can easily see the homogeneous dependence on $\frac {{\bar \mu}^2}
{m^2}$
by noting that if the nonpolynomial momentum dependent part has mass to
fulfill the homogeneity, then the coefficients must depend on mass in this
way. One can also see this point by directly solving the scale WT identity
in the agent constants version, i.e., using Eq. (12) or (5) that is valid
graph by graph.}. Thus we have locked the behavior of the underlying
structures acted by their agent(s). This is quite a natural conclusion
following from the standard underlying theory point of view. In a sense the
harmony notion elaborated here fixes the agent constants to a certain
degree. The fixation of the constants here corresponds to the subtraction
procedure in the conventional renormalization programs where the dependence
of the renormalized amplitudes upon the subtraction points can be almost
arbitrary. Our discussion above simply rejects all the schemes where the
dependence of the residual local parts of the elementary vertices upon the
subtraction scale is different from that given by Eq.(36) in the first place
before referring to other problems of the schemes. This could not have been
achieved in the conventional renormalization programs.

Therefore according to the foregoing discussions, the subtraction schemes
effected at Euclidean momentum are questionable for the massive case
\footnote{%
This is not to say the other schemes are totally OK. It is known that some
regularization schemes like dimensional regularization sometimes leads to
oversubtraction and are therefore questionable in certain applications \cite
{LENU}. The most crucial point is that all regularization schemes are ad hoc
ones contain various ill-definedness.}. Here we must mention that the
dependence on the agent constants in the way specified in Eq.(36) is
conventionally subject to the nondecoupling difficulty \cite{EFT}. However,

this is not a real challenge by noting that the underlying formulation is
correct only when the limit $L_{\{\sigma\}}$ is valid with respect to all
effective parameters including the agents. When a mass goes to infinity, the
expressions that have been obtained by assuming it is finite are no longer
true. We will make further discussions on this issue in section five.

The natural appearance of the dimensional agents in the way dictated in Eq.
(36) can also be technically understood in the following way. First we need
to briefly review the technique we proposed solely based on the existence of
the underlying structures for providing well-defined formulations \cite
{Usa} as follows: For a 1-loop graph $G$ of superficial divergence
degree $\omega_G-1$ \cite{Weinb} with its underlying version amplitude
denoted as $\Gamma_G([p],[g];\{\sigma\})$, we have 
\FL
\begin{eqnarray}
&&\partial^{\omega_G}_{[p]} L_{\{\sigma\}} \Gamma_G([p],[g]; \{\sigma\})=
L_{\{\sigma\}} \partial^{\omega_G}_{[p]} \Gamma_G([p],[g];\{\sigma\}) =
L_{\{\sigma\}} \int d^dk[\partial^{\omega_G}_{[p]} g([p,k],[g];\{\sigma\})]
\nonumber \\
&&=\int d^dk[\partial^{\omega_G}_{[p]}
L_{\{\sigma\}}g([p,k],[g];\{\sigma\})]=\int d^dk g^{\omega_G}
([p,k],[g];\{\sigma\})=\Gamma_G^{\omega_G}([p],[g]), \\
&&\Gamma_G([p],[g];\{c_G([g];\{\bar c\})\})= L_{\{\sigma\}}
\Gamma_G([p],[g];\{\sigma\}) \rightarrow \Gamma_G ([p],[g];\{c_G^u\})\equiv
[\partial^{\omega_G}_{[p]}]^{-1} \Gamma_G^{\omega_G}([p],[g])
\end{eqnarray}
where $\partial^{\omega_G}_{[p]}$ and its inverse denote the differentiation
(for $\omega$ times) with respect to the external momenta (one can do the
same with respect to masses in certain cases like for tadpole graphs) and
the indefinite integration respectively. We can not get the definite agent
constants or the coefficient constants but a set of unknown constants $%
\{c_G^u\}$ instead in our technical treatment. Eq.(37) follows from the
natural assumption that the underlying structures make the QFTs well defined
and the low energy limit does not alter the well-definedness.

Our above analysis implies that these unknown constants are to be defined as
the true agents through various physical properties. There must be at least
one constant with mass dimension among the constants $\{c_G^u\}$ as the
result of the indefinite integration though its precise value are to be
fixed otherwise and it must appear in the polynomial part of $\Gamma
_G([p],[g];\{c_G^u\})$ as $\sim \ln \frac{\mu ^2}{m^2}$ where $m$ is a mass
among the model constants $[g]$. This can be seen by observing that the
amplitude $\Gamma _G^{\omega _G}([p],[g])$ must take the following form
(with appropriate Feynman parametrization) 
\FL
\begin{equation}
\Gamma _G^{\omega _G}([p],[g])\sim \int_0^1[dx]f(\{x\})\frac{%
l([p],[g_m]|\{x\})}{Q([p],[g_m]|\{x\})}
\end{equation}
with $l(\cdots )$ and $Q(\cdots )$ denoting respectively the linear and
quadratic expressions of the combination of dimensional parameters like
external momenta and masses with the help of Feynman parameters $\{x\}$.
Then on the first step indefinite integration with respect to the external
parameters a logarithmic expression is generated with an arbitrary mass
scale entered for balancing the mass dimensions, 
\FL
\begin{equation}
\int_{indef}dp\Gamma _G^{\omega _G}([p],[g])\sim \int_0^1[dx]f(\{x\})\ln
\frac{Q([p],[g_m]|\{x\})}{\mu ^2}.
\end{equation}
Further indefinite integrations with respect to the external parameters of
mass dimension would necessarily carry on this mass scale in the polynomial
part. If one trades the $\mu $ in the logarithmic function with a model mass
constant so that the nonpolynomial function of momenta only involve model
mass parameters, the $\mu $ would only appear in the polynomial part in the
form of $\ln \frac{\mu ^2}{m^2}$ just as we claimed through scale anomaly
analysis. The multiplying factor of this logarithmic function is nothing
else but the ''anomalous'' dimension:$\delta _{\cdots }([g_0])$. Thus at
1-loop level, for all local ambiguities, we can claim that: 
\FL
\begin{equation}
c_{{\hat{O}}_{vert}}([g];\{{\bar{\mu}},{\bar{c}}_i^0\})=\frac 12\delta _{{%
\hat{O}}_{vert}}\ln
\frac{{\bar{\mu}}^2}{m^2}+C_{{\hat{O}}_{vert}}^0(\{{\bar{%
c}}_i^0\}).
\end{equation}
The superscript $0$ indicates that its principal is totally independent of
any constants with nonzero mass dimension.

Our technical proposal can be very naturally generalized to the multiloop
cases and many conventional subtleties like overlapping divergences are
automatically resolved, please refer to Ref. \cite{Usa} \footnote{%
Moreover, according to this technique, the new ambiguities in the
potentially divergent multiloop graphs beyond the subgraph ambiguities must
appear in the local part as $\ln \frac{\mu ^2}{m^2}$. Thus summing all the
overall local parts associated with each graph would lead to local vertices
with coefficients $\sim [1+\frac 12\delta _{O_v}([g^0])\ln \{\mu
^2/m^2\}+C_{O_v}^0(\{{\bar{c}}_i^0\})]$. That is the underlying structures
only ''renormalize'' the model constants $[g]$ by simple logarithmic
dependence on the masses and an agent scale.}. We will provide more examples
of calculations in the proposed technique in future reports, especially in
the nonperturbative case where the conventional regularization schemes often
yield more severe divergences.

Before closing this section we simply demonstrate that the asymptotic
freedom of QCD can be easily reproduced in our new version of CSE. The new
version CSE for the two-point connected correlation function of gauge fields
with external legs amputed reads (a specific gauge like Feynman gauge is
assumed) 
\FL
\begin{equation}
\{\lambda \partial _\lambda +{\bar{\delta}}_{g_s}g_s\partial _{g_s}+\sum
(1+{%
\bar{\delta}}_{m_i})m_i\partial _{m_i}-\delta _A\}D^c(\lambda
p,g_s,[m_i];\{{%
\bar{\mu}},c_j^0\})=0.
\end{equation}
For simplicity let us work in the massless case. Noting that
${\bar{\delta}}%
_{g_s}$ is equal to $-\frac 12\delta _{A_\mu }^{(1)}$ according to Eq. (16)
and rewriting the CSE in terms of $\alpha (=g_s^2/(4\pi ))$, we have ($%
\alpha _{eff}\equiv \alpha D^c$) 
\FL
\begin{equation}
\{\lambda \partial _\lambda -\delta _A\alpha \partial _\alpha -\delta
_A\}D^c(\lambda p,\alpha ;{\bar{\mu}})=0\Longrightarrow \{\lambda \partial
_\lambda -\delta _A\alpha \partial _\alpha \}\alpha _{eff}(\lambda p,\alpha
;%
{\bar{\mu}})=0.
\end{equation}
Here we have simply assumed that the constants independent of any mass scale
is zero or absorbed into the mass scale. Expanding $\alpha _{eff}$ in terms
of $\frac 12\ln [-p^2/{\bar{\mu}}^2]$ ($\alpha
_{eff}^{(1)}=\sum_{n=0}^\infty f_n(\alpha )t^n,t\equiv \frac 12\ln [-p^2/{%
\bar{\mu}}^2]$), one can find the following relation for the expansion
coefficients at one loop level of $\delta _A^{(1)}(=a\alpha )$ 
\FL
\begin{equation}
nf_n=\delta _A^{(1)}\alpha \partial _\alpha f_{n-1},\ \ f_0=\alpha ,\ \ \
\forall n\geq 1.
\end{equation}

The solution to Eq.(44) is 
\FL
\begin{equation}
f_n=a^n\alpha ^{n+1},\ \ \forall n\geq 0,\ \
\end{equation}
and hence the effective coupling takes the following well-known form
\FL
\begin{equation}
\alpha _{eff}^{(1)}(p^2,\alpha ;{\bar{\mu}})=\frac \alpha {1-\frac 12a\alpha
\ln \frac{-p^2}{{\bar{\mu}}^2}}.
\end{equation}
The asymptotic freedom is clearly reproduced for QCD due to the negative
constant $a$ at one-loop level. The IR Landau pole in the effective coupling
for QCD can not be trusted as we have worked in the UV end with IR details
coarse grained away. For QED, the Landau pole is UV one (due to positive
$a$%
), but it is not a final answer before the higher order corrections of $%
\delta _A$ are included. Moreover, as the energy is extremely higher, there
is another problem: the underlying structures would ''interfere with'' the
effective sectors more strongly, i.e., the present QFT models must be
modified somehow. This is closely related with the validity of the limit
operation $L_{\{\sigma \}}$ and will be discussed in more details in next
section. Hence the solution is trust worthy at energies not extremely high.

\section{Decoupling theorem and underlying structures}

Now let us turn to the decoupling issue of the heavy particles in QFTs.

As we have noted in last section, due to the logarithmic dependence of the
local constants ($c_{O_i}$) upon the particle mass ($\sim \ln m^2$,see
Eq.(41)) these constants would diverge as the particles become extremely
heavy ($M_h\equiv m\rightarrow \infty $) and this in turn implies the
nondecoupling of the heavy particles. While in the subtraction schemes
effected at Euclidean momentum, the heavy particles' contributions to the
light sector vanish as the masses go to infinity \cite{Appel,EFT}.

To resolve this scheme dependence of decoupling phenomenon, we note that
from the underlying theory point of view, when a physical mass or mass scale
goes to infinity relative to the characteristic energy scale of an effective
physical system, it means that the physics associated with this infinite
energy scale belongs to the UV underlying sectors of the effective system.
Or the heavy particles now join the underlying structures' party and the
underlying constants $\{\sigma \}$ are no longer infinities for such heavy
particles, then the expressions containing $\ln \frac{\mu ^2}{M_h^2}$ that
have arisen from the limit operation $L_{\{\sigma \}}$ is no longer valid,
or equivalently, the limit operation $\lim_{M_h\rightarrow \infty}$ can not
be directly applied to the Green functions simply because this operation
would inevitably invoke the underlying structures for it to make sense or
the following commutator is not zero, 
\FL
\begin{equation}
\left[ \lim_{M_h\rightarrow \infty},L_{\{\sigma \}}\right] \neq 0.
\end{equation}

So we must resort to the original underlying theory expressions which are
well defined no matter how heavy some fields (effective or fundamental) are.
Thus, when a field becomes too heavy to be excited by the low energy
interaction, or equivalently when the energy of a system becomes so low that
it crosses a heavy particle threshold, the underlying constants for the new
effective sectors now include the extremely heavy particles' mass. Then the
limit operation $L_{\{\sigma \}}$ now changes to $L_{\{\sigma ^{\prime }\}}$
with 
\FL
\begin{equation}
\left\{ \sigma ^{\prime }\right\} \equiv \left\{ \sigma \right\} \bigcup
\left\{ M_h\right\} =\left\{ \sigma ;M_h\right\} .
\end{equation}
These heavy particles' influences are delegated to the new agent constants
that finally appear from the new limit operation $L_{\{\sigma
^{\prime }\}}$%
. Since these agents are finite in principle, the effects of the heavy
particles only ''renormalize'' the light sectors by a finite amount in the
local vertices \cite{EFT}. Similarly, it is also easy to see that the
expressions obtained at finite momenta for QFTs can not be simply
extrapolated to the UV limit as it would again invalidate the limit
operation about the underlying structures. This is directly related to the
asymptotic expansion of operator products which will be discussed later in
our new formalism \cite{Y1}.

Equivalently, one can let the energy and masses of the light sectors become
relatively infinitesimal comparing to the heavy sectors, then the light
sectors would be again given in terms of masses and couplings belonging to
the light sectors and additional finite agent constants appearing only in
the local vertices of the light fields. This is just the spirit behind the
heavy quark field theory \cite{EFT}.

As the effective parameters like mass and momenta become extremely large,
such fields would interact more directly with the underlying structures.
Then the limit of decoupling the underlying structures for these heavy or
high-energy fields are no longer valid and one has to work in models
suitable for more higher energy physics. Thus the scheme dependence of the
decoupling theorem is again closely connected to the indispensable existence
of the underlying structures for making the QFTs naturally well defined. The
intimate relation between the decoupling theorem and the existence of the
underlying structures will be investigated intensively in the future.
Moreover, to make the formulation completely well defined, the IR underlying
structures should be included in all the future investigations.

Conventionally, the subtraction point is deemed as arbitrary, a running
scale. While in the underlying theory point of view, due to the condition of
decoupling theorem or the validity of the LE limit operation, such scales
(agent constants) should be uniquely defined up to possible equivalence. One
could not arbitrarily rescale any effective parameter (like mass and energy)
and any agent constant (which characterizes the quantum structures of an
effective sector in addition to the Planck constant and effective
parameters, see section II and III). Once a scale is extrapolated to a place
where the effective QFT models break down, the formulations in terms of the
LE effective parameters and agent constants cease to be correct or useful.

\section{Summary}

From our discussions above, we can see that the underlying theory postulate
is quite natural and powerful. Many subtle issues in conventional schemes
are resolved in a natural way in our proposal \cite{Usa}. The new equations
governing the scale behaviors are formulated in terms of canonical and agent
constants that are well defined. The most significant technical prediction
from our investigations here is the irrationality of certain conventional
renormalization schemes. We could also reproduce many novel results. The
derivation of trace anomaly in the underlying theory approach is a simple
work.

As the scheme dependence in high energy applications of QCD is a very
important issue, we hope our investigations here could be of help to this
issue. Of course our main goal here has been to demonstrate the plausibility
and power of our recent proposal for renormalization, further development
and applications of the proposal will be made in the future. We hope our
strategy could lead to more important contributions to high energy physics
and beyond. The issue of gauge dependence of the CSE in gauge field theories
will be discussed in the future together with the IR problem in such QFTs.

In summary, we derived several new versions of the Ward-Takahashi identities
of scale transformations in any QFT models along the natural strategy for
dealing with the UV ill-definedness. Among these new versions, each has its
own advantage and utility in different situations. Our expressions are all
free from divergences and the associated infinite renormalization. The new
equations improve the conventional ones in several aspects. A most important
new result is that the finite local part of any potentially divergent vertex
graph can not be arbitrary and therefore some conventional subtraction
schemes are, according to the underlying theory point of view, unreasonable
choices. The important decoupling theorem a la Applequist and Carazzone is
argued to be still valid outside Euclidean momenta subtraction schemes with
the underlying structures appropriately accounted.

\end{document}